\documentclass[prb,twocolumn,aps,showpacs,10pt,floatfix,longbibliography]{revtex4-1}

\usepackage[pdftex]{graphicx} % Include figure files
\usepackage{epstopdf}
\usepackage{verbatim}
\usepackage{color}
\usepackage{subfigure}
\usepackage{tabularx}
\usepackage{amsfonts}
\usepackage{amsmath}
\usepackage{wasysym}
\usepackage{bm}
\usepackage{hyperref}

\newcommand{\be}{\begin{equation}}
\newcommand{\ee}{\end{equation}}

\newcolumntype{C}[1]{>{\centering\let\newline\\\arraybackslash\hspace{0pt}}m{#1}}

\begin{document}
\title{Upper branch magnetism in quantum magnets: \\
Collapses of excited levels and emergent selection rules
}
\author{Changle Liu$^{1}$}
\thanks{These two authors contributed equally.}
\author{Fei-Ye Li$^{1}$}
\thanks{These two authors contributed equally.}
\author{Gang Chen$^{1,2}$}
\email{gangchen.physics@gmail.com}
\affiliation{$^{1}$State Key Laboratory of Surface Physics and Department of Physics, 
Fudan University, Shanghai 200433, China}
\affiliation{$^{2}$Department of Physics and Center of Theoretical and Computational Physics,
The University of Hong Kong, Pokfulam Road, Hong Kong, China}

\date{\today}

\begin{abstract}
In many quantum magnets especially the rare-earth ones, the low-lying crystal field 
states are not well separated from the excited ones and thus are insufficient to describe 
the low-temperature magnetic properties. Inspired by this simple observation, 
we develop a microscopic theory to 
describe the magnetic physics due to the collapses of the weak crystal field states. We 
find two cases where the excited crystal field states should be seriously included 
into the theory. One case is when the bandwidth of the excited crystal field states 
is comparable to the crystal field gap. The other case is when the exchange 
energy gain between the low-lying and excited crystal field states overcomes 
the crystal field gap. Both cases could drive a phase transition and result 
in magnetic orders by involving the excited crystal field states. We dub the 
above physics as upper branch magnetism and phase transition. 
We discuss the multitude of magnetic phases and the emergent selection rules
for the detection of the underlying excitations. We expect our results 
to help improve the understanding of many rare-earth magnets with weak 
crystal field gaps such as Tb$_2$Ti$_2$O$_7$ and Tb$_2$Sn$_2$O$_7$,
and also provide a complementary perspective to the prevailing 
local ``$J$'' physics
in $4d$/$5d$ magnets.
\end{abstract}

\maketitle

\section{Introduction}

Frustrated quantum magnetism has been a rather active 
field of research, both theoretically and 
experimentally~\cite{balents2010spin,doi:10.1139/p01-123}. 
The interest not only
lies on the possibility of searching for novel quantum phase of matter and 
the related phenomena~\cite{balents2010spin,doi:10.1139/p01-123,0034-4885-80-1-016502,palee,RevModPhys.88.041002}, 
but also arises from the vast families of quantum materials with frustrated magnetic 
interactions~\cite{Gingras2014,balents2010spin,0034-4885-80-1-016502,RevModPhys.82.53,RauGingras}. 
This prosperous field requires both the microscopic understanding of 
the quantum materials and their quantum chemistry and the theoretical understanding 
of abstract and fundamental concepts for quantum matter, and more importantly,  
establishing the bridge between fundamental theories and experimental phenomena. 
The first step in the theoretical understanding of these complex quantum magnets 
is to understand the microscopic degrees of freedom. For (magnetic) Mott insulators, 
the conventional recipe is based on the Hund's rules that clarify the microscopic 
local moment structure~\cite{patrik1999lecture}. From the interaction of the local moments, one can 
then establish a microscopic many-body model for the system.

For the rare-earth based quantum magnets that
are currently under an active study~\cite{Gingras2014,0034-4885-80-1-016502,RevModPhys.82.53}, 
the crystal field effect is an important ingredient
in the understanding of the microscopics. The widely accepted standard approach 
is to understand the crystal field level and find the local ground states~\cite{patrik1999lecture}. 
The local ground states can be usual Kramers doublets~\cite{PhysRevX.1.021002,PhysRevLett.106.187202,PhysRevB.88.134428,YueshengPRL1,PhysRevB.94.035107,PhysRevB.98.054408,PhysRevLett.119.127201,PhysRevB.93.104405,onoda2011effective}, 
dipole-octupole doublets~\cite{huang2014quantum,PhysRevB.95.041106,PhysRevB.94.201114,PhysRevLett.115.197202,PhysRevB.94.104430,PhysRevB.92.224430,fragmentation},
and non-Kramers doublets~\cite{PhysRevLett.105.047201,PhysRevB.83.094411,PhysRevB.86.104412,PhysRevB.96.195127,PhysRevB.94.205107,PhysRevLett.118.107206,PhysRevB.92.054432,GCnonK}, 
and these local ground states control the low-temperature
magnetic properties of many rare-earth magnets. This standard approach has 
found a great success in the study of rare-earth pyrochlore magnets such as  
Yb$_2$Ti$_2$O$_7$~\cite{PhysRevX.1.021002,PhysRevLett.119.057203,PhysRevB.98.054408,PhysRevB.95.094422}
and Er$_2$Ti$_2$O$_7$~\cite{PhysRevB.95.094422,PhysRevLett.109.167201,PhysRevLett.109.077204,PhysRevLett.112.057201,PhysRevB.93.184408}
rare-earth triangular magnets such as YbMgGaO$_4$~\cite{YueshengPRL1,YueshengSR,YMGOYao,PhysRevX.8.031001,PhysRevB.96.054445,PhysRevB.97.125105,PhysRevLett.117.267202,nphys2017,PhysRevLett.119.157201,PhysRevLett.120.207203,Maksimov,YShenNcomm,PhysRevX.8.031028,PhysRevLett.118.107202,PhysRevLett.117.097201}, 
and rare-earth based double perovskites~\cite{PhysRevB.95.085132}. 
The local moment structure of the iridate family  
such as hyperkagome Na$_4$Ir$_3$O$_8$~\cite{PhysRevLett.99.137207,PhysRevLett.100.227201,PhysRevB.78.094403}, honeycomb Na$_2$IrO$_3$~\cite{PhysRevLett.102.017205,PhysRevLett.105.027204,PhysRevLett.110.097204},
hyper-honeycomb $\beta$-Li$_2$IrO$_3$~\cite{PhysRevLett.114.077202}, harmonic-honeycomb 
Li$_2$IrO$_3$~\cite{Analytis}
may also be interpreted under this scheme. 
The success of this approach requires that the crystal field gap between
the ground state doublet and the excited one is sufficiently large. 
This requirement, however, may not be always satisfied in many systems. 
In this paper, we provide a non-standard approach to understand the 
magnetic physics for the systems when the crystal field gap is not so 
large. For our purpose, we include the excited crystal field levels
and consider the interactions between these levels from different lattice 
sites. At the same time, we consider the hybridization between the 
excited levels and the ground state level from the neighboring sites. 
From these new ingredients in the microscopic analysis, we illustrate 
this new theoretical framework by applying to a specific example
on a face centered cubic (FCC) lattice. Because the magnetic 
physics emerges from the excited energy levels, 
we dub this piece of physics as the upper branch magnetism. 
%The notion of ``upper branch'' originates from the context of ultracold atoms, molecules and optical physics where the cold atoms on the optical lattice can occupy the excited$p$ and $d$ bands/orbitals and these metastable states contribute to the physics. We think the usage of ``upper branch'' should be appropriate for our magnetic context here. 

The remaining part of the paper is organized as follows. In Sec.~\ref{sec2}, 
we first introduce our microscopic model for the specific case that we consider. 
In Sec.~\ref{sec3}, we use both the Weiss mean-field method and the flavor wave theory to establish the full phase 
diagram of this model and explain the magnetic excitations. 
In Sec.~\ref{sec4}, we explain the emergent selection rules for the detection 
of magnetic excitations. This is associated with certain symmetry properties 
of the ground state wavefunctions of the relevant magnetic phases.
In Sec.~\ref{sec5}, we conclude with a discussion about the general applicability 
of our understanding to weak crystal field quantum magnets 
and present a list of systems and materials that share some similarities in terms 
of local energy level schemes, phase transitions and universality.

\section{Microscopic model}
\label{sec2}

\begin{figure}[t]
\includegraphics[width=6.5cm]{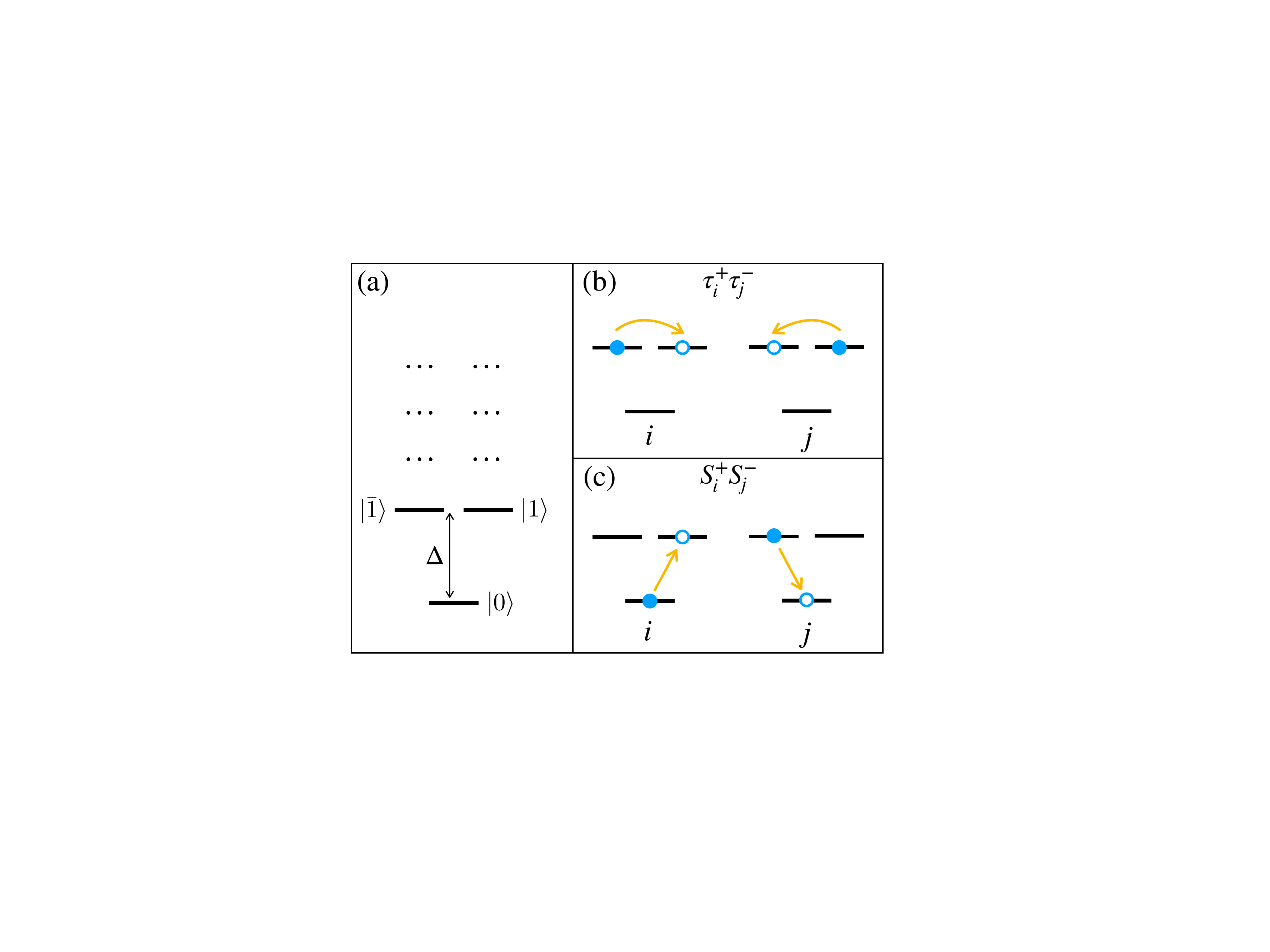}
\caption{(a) The crystal field level scheme for a magnetic ion on a lattice. 
The lowest three states are viewed as an effective spin-1 local moment. 
``$\Delta$'' is the crystal field gap. ``$\cdots$'' refers to 
the excited doublets that are neglected in our theoretical analysis. 
(b) The $\tau_i^+\tau_j^-$ process.
(c) One possible $S_i^+S_j^-$ process.
} 
\label{fig1}
\end{figure}

The magnetic physics out of the weak crystal 
field levels is quite general and applies to many different systems with 
different lattice geometries. To illustrate the essential physics, we merely focus 
on one lattice. We study the interacting local moments with weak crystal
fields on the FCC lattice. To further simplify the model and keep the essential
physics, we consider the local crystal field level scheme in Fig.~\ref{fig1}. 
The local ground state here is a singlet, and the first excited state is 
a doublet. This crystal field scheme could occur for the rare-earth ion 
with even number of electrons~\cite{0953-8984-24-25-256003}. For the rare-earth ion 
with odd number of electrons, the ground state must at least be a doublet (though the possibility of being a quartet may still remain),
and one then needs to consider the interaction between these doublets. 
This would complicate the problem and cover the essential physics that 
is uncovered in this work.

If the crystal field gap, $\Delta$, is much larger than other energy scales 
that are specified below, the ground state of the system would be a trivial 
product state of the local singlets. There are other microscopic processes
that compete with the crystal field gap. As we show in Fig.~\ref{fig1},  
one process is the superexchange interaction between the upper doublets. 
The other process is the hybridization between the ground state and 
the excited doublet. Fundamentally, both processes arise from the superexchange 
interactions. To distinguish them, however, we quote them differently.

To model the minimal microscopic physics, we neglect the further excited states 
beyond the three states in Fig.~\ref{fig1}. The three states, one ground state
singlet and one excited doublet, can be thought as an effective spin-1 local 
moment. We identify the ground state singlet as the ${S^z=0}$ state, and identify 
the excited doublet as ${S^z =\pm 1}$ states. From this mapping, the crystal field
splitting can be regarded as a single-ion anisotropy, {\sl i.e.},
\begin{eqnarray}
H_{\text{CEF}} =  \Delta \sum_i  (S_i^z)^2 . 
\end{eqnarray}

The superexchange interaction between the upper doublet is given as 
\begin{eqnarray}
H_{\text{ex}} &=& \sum_{\langle ij \rangle} J_z \tau^z_i \tau^z_j - J_{\perp} 
(\tau^+_i \tau^-_j + h.c.) + \cdots ,
\end{eqnarray}
where the pseudospin-1/2 operator $\boldsymbol{\tau}_i$ operates 
on the upper doublet, and ``$\cdots$'' refers to other interactions
such as $\tau^+_i \tau^+_j$ and $\tau^-_i \tau^-_j$. 
It is straight-forward to establish the following relation between the 
pseudospins and the spin-1 operators,
\begin{eqnarray}
\tau^z_i &\equiv & \frac{1}{2} P_i S^z_i P_i,  \\ 
\tau^{\pm}_i & \equiv & \frac{1}{2} P_i (S_i^{\pm})^2 P_i, 
\end{eqnarray}
where $P_i$ is a projection operator onto the upper doublet.

We introduce the hybridization as a conventional superexchange,
{\sl i.e.},
\begin{eqnarray}
H_{\text{hybrid}} = - \sum_{\langle ij \rangle} 
                    J_{h} ( S^+_i S^-_j  + S^-_i S^+_j ). 
\end{eqnarray}

Summarizing the above results, we have our full minimal model 
as ${H= H_{\text{CEF}} + H_{\text{ex}} + H_{\text{hybrid}}}$. 
This model hosts an $U(1)$ symmetry generated by spin rotation of 
arbitrary angle about the $z$ axis. This continuous symmetry is due
to oversimplification of our model, and is expected to vanish for
realistic materials with strong spin-orbit coupling.

For the most common cases where the crystal field has $D_{3d}$ point
group symmetry, all effective spin components $S^{\mu}$ ($\mu=x,y,z$)
are time-reversal odd and behave as dipole moments, while the spin
bilinear terms behave as quadrupoles. Thus $\boldsymbol{S}$
and $\tau^{\pm}$ can be served as dipolar and quadrupolar order parameters, 
respectively. It should be noticed that as the crystal field splitting
term $(S^{z})^{2}$ explicitly enters the Hamiltonian, the quadrupole component
$Q_{3z^2-r^2}\equiv3(S^z)^2-2$ must have non-zero expectation value inside
each phase. It is preformed and can not be served as an order parameter 
related to a symmetry breaking.

\begin{table*}[t]
	\begin{tabular}{ccccc}
		\hline\hline
		Magnetic phases & Dipoles & Quadrupoles & Local variational wavefunction & Variational energy per site
		\\
		\hline
		Quantum paramagnet & $ \langle \boldsymbol{S} \rangle = 0 $ & $ \langle {\tau^\pm} \rangle = 0 $ 
		&  $ |0\rangle $ & $ 0 $
		\\
		FM$_\text{xy}$  & $ \langle {S^{\pm}} \rangle \neq 0 $ & $ \langle {\tau^\pm} \rangle = 0 $
		& $ c\, e^{i \theta} |1\rangle + \sqrt{1 - 2c^2}\, |0\rangle +  c\, e^{-i \theta} |\bar{1}\rangle $ 
			& $ -192c^4 J_h $ 
		\\
		FM$_\text{z}$  & $ \langle {S^{z}} \rangle \neq 0 $ & $ \langle {\tau^\pm} \rangle = 0 $
		& $ |1\rangle $ or $|\bar{1}\rangle $ & $ \frac{3J_z}{2} + \Delta $
		\\
		AFM$_\text{z}$  & $ \langle {S^{z}} \rangle \neq 0 $ & $ \langle {\tau^\pm} \rangle = 0 $
		& $ |e^{i \boldsymbol{Q} \cdot \boldsymbol{r}_i}\rangle $ with $ \boldsymbol{Q}=(0,0,2\pi) $
			& $ -\frac{J_z}{2} + \Delta $
		\\
		FQ$_\text{xy}$  & $ \langle \boldsymbol{S} \rangle = 0 $ & $ \langle {\tau^\pm} \rangle \neq 0 $
		&  $ \frac{1}{\sqrt{2}} \big(e^{i\theta/2}|1\rangle+e^{-i\theta/2}|\bar{1}\rangle\big) $
			& $ -3J_\perp + \Delta $
		\\
		\hline\hline
	\end{tabular}
	\caption{Physical properties of different magnetic phases. 	
	       Both the FM$_{\text{xy}}$ state and the FQ$_{\text{xy}}$ state 
			have an $U(1)$ degeneracy characterized by the angular variable $\theta$.
			For the FM$_{\text{xy}}$ state, $c^2\equiv({48 J_h - \Delta})/({192 J_h})$. 
			Note that when $J_h/\Delta=1/48$ we have $c=0$, $|\Psi\rangle_i=|0\rangle$,
			and when $J_h\rightarrow\infty$ we have $c=1/2$,
			$|\Psi\rangle_i$ becomes fully polarized in the $xy$ plane.}
	\label{tab1}
\end{table*}

\section{Phase diagram}
\label{sec3}

\begin{figure}[t]
	\hspace{-5mm}
 	\includegraphics[width=6.8cm]{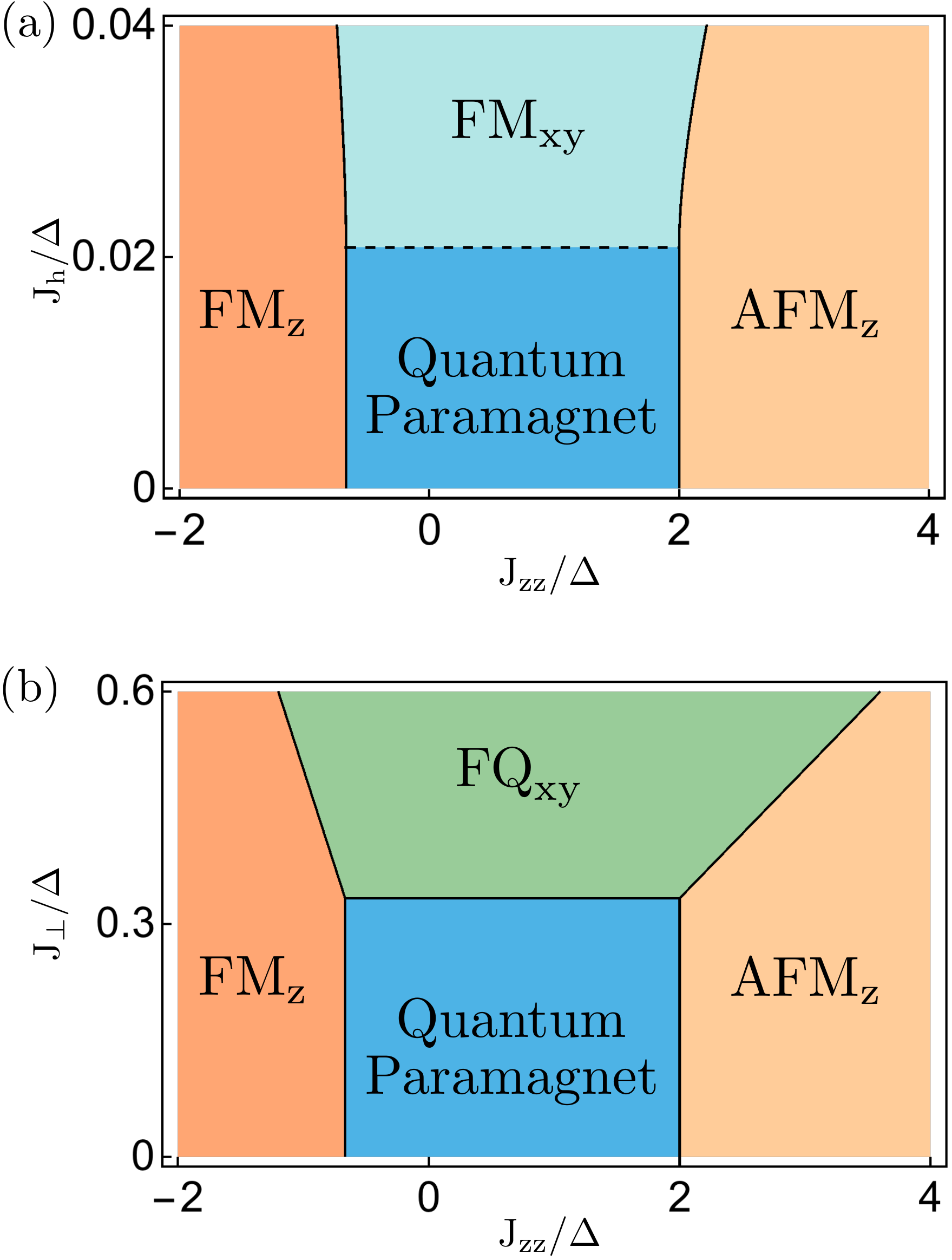}
	\caption{(Color online.) Phase diagram for 
	    (a) ${J_{\perp}=0}$ and    
	    (b) ${J_{h}=0}$.
		The details of the ordered phases are explained in the main text.
		The dashed line in (a) refers to a continuous phase transition and
		the remaining transitions are first order.
	}
\label{fig2}
\end{figure}

\subsection{Weiss mean-field theory}

To establish the ground state phase diagram of the full model, 
we adopt the Weiss mean-field method to decouple interactions
between different sites. Here we set up a two-sublattice ansatz 
that is consistent with the ${\boldsymbol{Q}=(0,0,2\pi)}$ antiferromagnetic 
configuration on the FCC lattice. This type of antiferromagnetic 
ordering is a common order pattern on an antiferromagnetically   
interacting FCC magnet. 
The Weiss mean-field approach is essentially a variational approach  
with simple product states as the variational wavefunction.
We here express the trial ground state wavefunction as a product state 
\begin{equation}
	|\Psi_\text{GS}\rangle=\prod_{i} |\Psi\rangle_i.
\end{equation}
The single-site state $|\Psi\rangle_i$ is then determined by minimizing
the energy per site $E=\langle\Psi_\text{GS}| H |\Psi_\text{GS}\rangle / N$.
The results are listed in Table~\ref{tab1} using the $S^z$ diagonal basis 
$|S^z=1,0,\bar{1}\rangle$. 
Our mean-field phase diagram is shown in Fig.~\ref{fig2}.  
We can understand the magnetic phases with interactions involving 
upper branch (excited) states. The corresponding order parameters 
for different phases are listed in Table~\ref{tab1}.

In the large $\Delta$ limit where the crystal field gap is dominant,  
the ground state is a trivial product state of non-magnetic singlet 
on each lattice site, which we dub as the quantum paramagnetic state. 
Such a state is protected by the energy gap $\Delta$ and hence is 
stable against small perturbations. When the interactions in $H_{\text{ex}}$ 
and $H_{\text{hybrid}}$ become large enough such that the energy      
gain from the exchange or hybridization overcomes the crystal field gap, 
various magnetic orders can be realized. The $J_{z}$ term favors spins 
to form a dipolar order along the $z$ axis, leading to the FM$_{\text z}$ 
(AFM$_{\text z}$) state in the left (right) side of the quantum paramagnetic 
state (see Fig.~\ref{fig2}(a) and (b)). Similarly, the $J_{\perp}$ term 
favors the dipolar order with spins on the $xy$ plane.
This gives the FM$_{\text{xy}}$ state in the upper side of Fig.~\ref{fig2}(a).
Finally, the hybridization term $J_{h}$ term favors spins to form quadrupolar 
order with the director of the quadrupolar order on the $xy$
plane, giving the FQ$_{\text{xy}}$ state in the upper side of Fig.~\ref{fig2}(b).
Both the FM$_{\text{xy}}$ and FQ$_{\text{xy}}$ state spontaneously 
break the global $U(1)$ symmetry. 
The details of these states can be found in Table.~\ref{tab1}.

Although both $H_{\text{ex}}$ and $H_{\text{hybrid}}$ could drive the 
system out of the quantum paramagnetic state, the mechanisms 
in which the excited levels are involved are quite different.
For the former case, the neighboring magnetic states are driven 
by exchange terms through first-order transitions, reflecting 
the competition between the crystal field splitting and the exchange 
energy gain purely within the excited levels through, for example, 
the $\tau_i^+\tau_j^-$ process shown in Fig.~\ref{fig1}(b). 
For the latter case, however, the hybridization term drives 
the system into the FQ$_{\text{xy}}$ state through a continuous transition. 
The excited states can hop between different sites via the hybridization
processes and one of such processes is shown in Fig.~\ref{fig1}(c).
$H_{\text{hybrid}}$ hence introduce bandwidth to the excited states 
and the criticality is obtained when the bandwidth is
comparable with the crystal field gap. To further clarify this point, 
we study the instability of the quantum paramagnetic state from the 
flavor wave theory below.

\subsection{Flavor wave theory and magnetic excitations}

We adopt the flavor wave theory to investigate the spin excitations and reveal 
the magnetic instability of the quantum paramagnetic state~\cite{PhysRevB.60.6584}.
Within this framework, different roles that are played by $H_{\text{ex}}$ 
and $H_{\text{hybrid}}$ will be clear.

Under the flavor wave representation, the internal states of a spin 
at site $i$ are represented by three flavors of bosons,
\begin{equation}
b_{im}^{\dagger}|\Omega\rangle\equiv|S_{i}^{z}=m\rangle,
\end{equation}
and arbitrary onsite spin operator $O_i$ can be written as 
${\sum_{m,n} \langle m|O|n \rangle b_{im}^{\dagger}b_{in}^{\phantom{\dagger}}}$
with ${m,n=1,0,\bar{1}}$. The physical Hilbert space is recovered under the local 
Hilbert space constraint ${\sum_{m} b_{im}^{\dagger}b_{im}^{\phantom{\dagger}}=1}$.

In the following we will omit the site index $i$ for simplicity.
The relevant onsite spin operators can be written as
\begin{eqnarray}
&& \tau^{z}  = \frac{1}{2}\big(b_{1}^{\dagger}b_{1}^{\phantom{\dagger}}-b_{\bar{1}}^{\dagger}b_{\bar{1}}^{\phantom{\dagger}}\big),\\
&& \tau^{+}  = b_{1}^{\dagger}b_{\bar{1}}^{\phantom{\dagger}},\,\,\,
\tau^{-}  = b_{\bar{1}}^{\dagger}b_{1}^{\phantom{\dagger}},\\
&& (S^{z})^2 = b_{1}^{\dagger}b_{1}^{\phantom{\dagger}}+b_{\bar{1}}^{\dagger}b_{\bar{1}}^{\phantom{\dagger}},\\
&& S^{+} = \sqrt{2}\big(b_{1}^{\dagger}b_{0}^{\phantom{\dagger}}+b_{0}^{\dagger}b_{\bar{1}}^{\phantom{\dagger}}\big),\\
&& S^{-} = \sqrt{2}\big(b_{\bar{1}}^{\dagger}b_{0}^{\phantom{\dagger}}+b_{0}^{\dagger}b_{1}^{\phantom{\dagger}}\big).
\end{eqnarray}

In the language of the flavor wave theory, different magnetic phases 
can be obtained by condensing corresponding flavors of the bosons. 
For the quantum paramagnet of this subsection, 
the $b_0$ flavor is condensed
and
\begin{equation}
b_{i0}^{\dagger}=b_{i0}^{\phantom \dagger}
\simeq
[1-b_{i1}^{\dagger}b_{i1}^{\phantom \dagger}-b_{i\bar{1}}^{\dagger} 
b_{i\bar{1}}^{\phantom \dagger} ]^{1/2},
\end{equation}
while the other two flavors of bosons represent the excitated states
above the quantum paramagnetic state. The (quadratic) flavor wave 
Hamiltonian is then obtained as
\begin{equation}
H=\sum_{\boldsymbol{k}} \Psi_{\boldsymbol{k}}^{\dagger} 
\begin{pmatrix}B_{\boldsymbol{k}}+\Delta & B_{\boldsymbol{k}}\\
B_{\boldsymbol{k}} & B_{\boldsymbol{k}}+\Delta
\end{pmatrix}
\Psi_{\boldsymbol{k}},
\end{equation}
where 
${\Psi_{\boldsymbol{k}}=(b_{\boldsymbol{k}1}^{\phantom \dagger}, 
b_{-{\boldsymbol{k}}\bar{1}}^{\dagger})^T}$
and ${B_{\boldsymbol{k}}=-2J_h\gamma_{\boldsymbol{k}}}$,
with $\gamma_{\boldsymbol{k}}=\sum_{\boldsymbol{\delta}}
\cos (\boldsymbol{k}\cdot{\boldsymbol{\delta}} )$
and ${\boldsymbol{\delta}}$ is summed over the twelve nearest neighbor 
vectors of the FCC lattice.
The magnetic excitation with respect of the quantum paramagnetic state 
has a two-fold degeneracy that is protected by the time-reversal symmetry 
of the quantum paramagnet. The dispersion of the magnetic excitations is 
\begin{equation}
	\omega_{\boldsymbol{k}}=[{\Delta(2B_{\boldsymbol{k}}+\Delta)}]^{1/2}.
\end{equation}
The magnetic excitations are fully gapped inside the quantum paramagnetic 
state. As the system approaches the transition to a proximate ordered 
phase, the gap of the excitation is closed. The closing point is at the 
$\Gamma$ point and ${J_h=\Delta/48}$. At this critical point, the excitation 
spectrum disperses linearly near the $\Gamma$ point, 
contributing to a ${C_{v}\sim T^{3}}$ heat capacity 
behavior at low temperatures.

It is apparent that the (quadratic) flavor wave Hamiltonian depends on $J_{h}$ 
while it does not depend on $J_{z}$ or $J_{\perp}$. As we have previously 
discussed, the hybridization term $J_{h}$ would create quantum fluctuations above 
the quantum paramagnetic state and bring the dispersion to the excitation 
crystal field state. However, the exchange part $H_{\text{ex}}$ only acts 
within the upper branch excited states and does not mix 
the upper branches with the local crystal field ground state. 
Therefore, the induced transitions from the quantum paramagnet to ordered
states through $H_{\text{ex}}$ must be strongly first order.

\subsection{Flavor wave theory for FM$_{\text{xy}}$ state}
\label{sec3c}

For the FM$_{\text{xy}}$ state, we choose the magnetization along the $\hat{x}$ direction
to break the continuous U(1) symmetry such that the variational wavefunction has the form 
as 
\begin{equation}
|\Psi\rangle_i=c\, |1\rangle + \sqrt{1 - 2c^2}\, |0\rangle +  c\, |\bar{1}\rangle.
\end{equation}
We introduce a new basis for the three favors of bosons via an unitary transformation,
\begin{equation}
\begin{pmatrix}  
a_0 \\ a_1 \\ a_2
\end{pmatrix}
= 
\begin{pmatrix}  
c & \sqrt{1 - 2c^2} & c
\\ 
\frac{1}{\sqrt{2}} & 0 & -\frac{1}{\sqrt{2}}
\\ 
\sqrt{\frac{1-2c^2}{2}} & -{\sqrt{2}} c & \sqrt{\frac{1-2c^2}{2}}
\end{pmatrix}
\begin{pmatrix}  
b_1 \\ b_0 \\ b_{\bar{1}}
\end{pmatrix},
\end{equation}
and condense the $a_0$ flavor. The quadratic flavor wave Hamiltonian reads
	\begin{equation}
	H=\frac{1}{2} \sum_{\boldsymbol{k}} \Phi_{\boldsymbol{k}}^{\dagger} 
	\begin{pmatrix}
	m_{11} & 0 & m_{13} & 0 \\
	0 & m_{22} & 0 & m_{24} \\
	m_{13} & 0 & m_{11} & 0 \\
	0 & m_{24} & 0 & m_{22}
	\end{pmatrix}
	\Phi_{\boldsymbol{k}},
	\end{equation}
where the entries in the matrix are given by  
\begin{align}
m_{11}=&(1-2c^2)(192c^2J_h+\Delta) \nonumber \\
& - \big[4 (1-2c^2) J_h - c^2 J_z\big]\gamma_{\boldsymbol{k}}/2, \\
m_{13}=&-\big[4(1-2c^2) J_h + c^2 J_z\big] \gamma_{\boldsymbol{k}} /2, \\
m_{22}=&(1-2c^2)(192c^2J_h+\Delta)-\Delta   \nonumber \\
&-2(1-4c^2)^2 J_h \gamma_{\boldsymbol{k}}, \\
m_{24}=& 2(1-4c^2)^2 J_h \gamma_{\boldsymbol{k}},
\end{align}
and ${\Phi_{\boldsymbol{k}} \equiv
\big(a_{\boldsymbol{k}{1}}^{\phantom\dagger},
a_{\boldsymbol{k}2}^{\phantom\dagger},
a_{\boldsymbol{-k}\bar{1}}^{\dagger},
a_{\boldsymbol{-k}\bar{2}}^{\dagger}\big)}$.

\begin{figure*}[t]
\includegraphics[width=\textwidth]{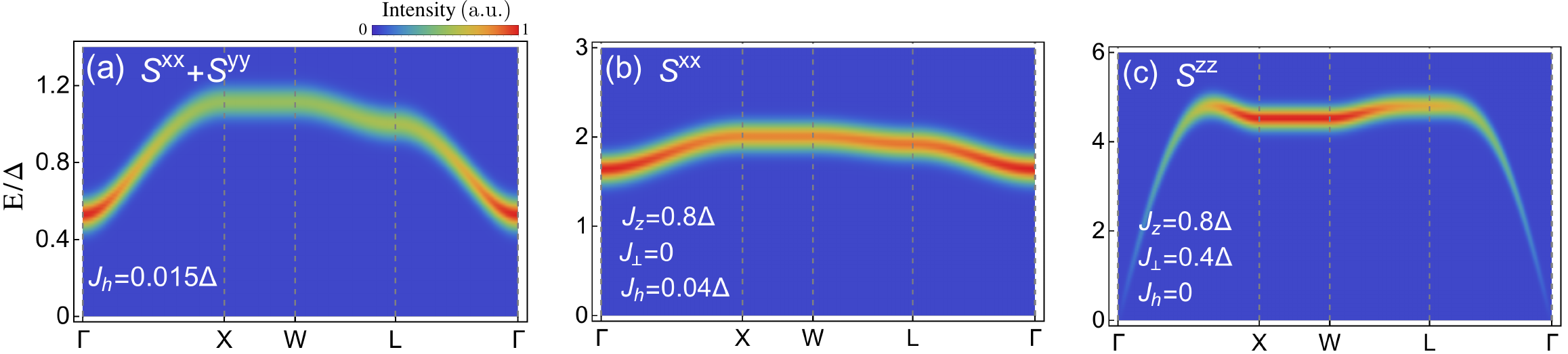}
	\caption{(Color online.) Dynamic spin structure factors of (a) the quantum paramagnet, 
	   (b) the FM$_\text{xy}$ state, and (c) the FQ$_\text{xy}$ state. 
		All the three states have non-vanishing quadrupole components that
		can be detected via inelastic neutron scattering measurements with
		corresponding polarization channels.
		The high symmetry points in the Brillouin zone are defined as
		${\Gamma=(0,0,0)}$, ${X=(0,2\pi,0)}$, ${W=(\pi,2\pi,0)}$, ${L=(\pi,\pi,\pi)}$.
		We set the magnetization of the FM$_\text{xy}$ state along the $\hat{x}$ direction,
		and the director of the quadrupolar order in the FQ$_\text{xy}$ state parallel to the $x$ axis.		
		Only coherent excitations are considered here. 
		The full spectra of the flavor wave 
		excitations are shown in Fig.~\ref{fig4} as a comparison.   
	}
	\label{fig3} 
\end{figure*}

\subsection{Flavor wave theory for FQ$_{\text{xy}}$ state}
\label{sec3d}

For the FQ$_{\text{xy}}$ state, we choose the director of 
this quadrupolar state along the $x$ axis to break the 
U(1) rotational symmetry.  
The single-site variational wave function is given as
\begin{equation}
|\Psi\rangle_i=\frac{i}{\sqrt{2}}\big(|1\rangle-|\bar{1}\rangle\big).
\end{equation}
To describe the elementary excitations with respect to this ground state,
we introduce a new basis for the flavors of bosons by 
the following transformation 
\begin{equation}
\begin{pmatrix}  
a_0 \\ a_1 \\ a_2
\end{pmatrix}
= 
\begin{pmatrix}  
\frac{-i}{\sqrt{2}} & 0 & \frac{i}{\sqrt{2}}
\\ 
\frac{1}{\sqrt{2}} & 0 & \frac{1}{\sqrt{2}}
\\ 
0 & i & 0
\end{pmatrix}
\begin{pmatrix}  
b_1 \\ b_0 \\ b_{\bar{1}}
\end{pmatrix},
\end{equation}
and condense the $a_0$ flavor. The quadratic flavor wave Hamiltonian reads
	\begin{equation}
	H=\frac{1}{2} \sum_{\boldsymbol{k}} \Phi_{\boldsymbol{k}}^{\dagger} 
	\begin{pmatrix}
	m_{11} & 0 & m_{13} & 0 \\
	0 & m_{22} & 0 & m_{24} \\
	m_{13} & 0 & m_{11} & 0 \\
	0 & m_{24} & 0 & m_{22}
	\end{pmatrix}
	\Phi_{\boldsymbol{k}},
	\end{equation}
	where the matrix entries are given as  
	\begin{align}
	m_{11}&=12 J_{\perp} - \gamma_{\boldsymbol{k}} (2J_{\perp} - J_z)/4, \\
	m_{13}&=-\gamma_{\boldsymbol{k}}(2J_{\perp} + J_z)/4, \\
	m_{22}&=6 J_{\perp} - \Delta, \\
	m_{24}&=0,
	\end{align}
	and $\Phi_{\boldsymbol{k}} \equiv
	\big(a_{\boldsymbol{k}{1}}^{\phantom\dagger},a_{\boldsymbol{k}2}^{\phantom\dagger},
	a_{\boldsymbol{-k}\bar{1}}^{\dagger},a_{\boldsymbol{-k}\bar{2}}^{\dagger}\big)$.

\section{Emergent selection rules}
\label{sec4}

\subsection{Selective measurements of dipole and quadrupole moments}

\begin{figure*}[t]
 	\includegraphics[width=\textwidth]{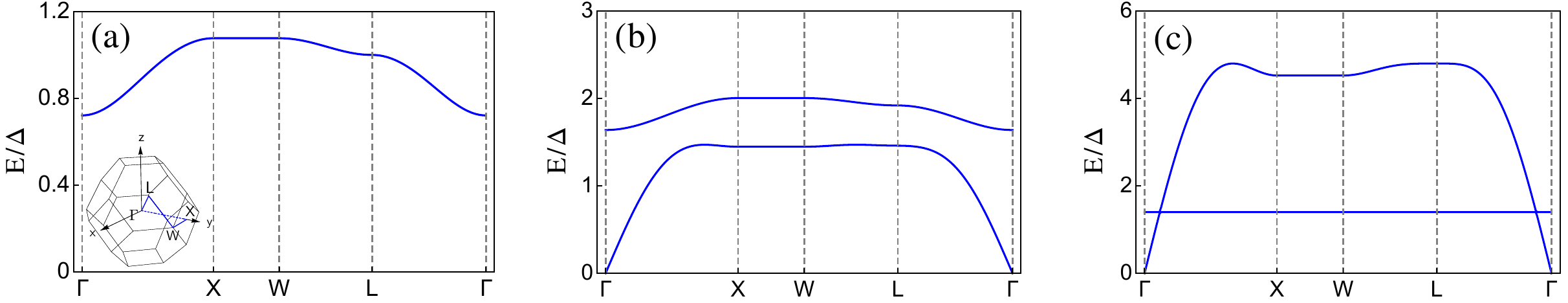}
	\caption{Full flavor wave excitations for (a) the quantum paramagnet, 
	   (b) the FM$_\text{xy}$ state, and (c) the FQ$_\text{xy}$ state, with the same 
	   parameters as chosen in Fig.~\ref{fig3} in the main text.
		The inset in (a) indicates the high symmetry points in the Brillouin zone of the FCC lattice.
	}
	\label{fig4}
\end{figure*}

In conventional experimental measurements such as neutron and $\mu$SR, 
one can only probe the dipolar orders while the quadrupolar orders are 
not directly visible in conventional magnetic measurements. 
The question thus arises that how to experimentally detect the 
phases accompanied by quadrupole components in our phase diagram.
In previous works some of the authors have proposed the scheme of 
selective measurements of dipole and quadrupole components using
elastic and inelastic probes~\cite{PhysRevB.94.201114,GCnonK,shen2018hidden}, 
and similar approach can be used here. The dipole moment, 
$\boldsymbol{S}$, is directly visible through conventional
magnetic measurements such as NMR and elastic neutron experiment.
Since the quadrupole moments do not commute with dipolar ones along orthogonal
directions, when the neutron scattering measures the dipole components, it 
creates quantum fluctuations to the orthogonal quadrupoles, leading to coherent
spin-wave-like excitations. This means that although the quadrupole itself
is invisible in an usual neutron scatterng measurement, 
the dynamic excitations of quadrupolar orders can be
visible in experiments and these excitations carry information of 
the underlying quadrupole structure.

To demonstrate the above discussion, here we calculate the dynamic spin 
structure factors for three representative states in Fig.~\ref{fig2} 
using the flavor wave theory. The results are shown in Fig.~\ref{fig3}.
In the polarized inelastic neutron scattering
experiment one measures the dynamic spin structure factors
\begin{eqnarray}
&& {\mathcal{S}}^{\mu\nu}(\boldsymbol{k},\omega) \nonumber \\
&& =\frac{1}{2\pi N}
\sum_{ij}\int_{-\infty}^{+\infty}\mbox{d}t\,e^{i\boldsymbol{k}\cdot(\mathbf{r}_{i}-\mathbf{r}_{j})-i\omega t}
\langle S_{i}^{\mu}(0)S_{j}^{\nu}(t)\rangle,
\end{eqnarray}
where $\mu,\nu=x,y,z$ represent the polarizations of incoming and
scattered neutrons. Thus one can read off signatures of the dipole 
and quadrupole components separately from the elastic and inelastic probes.

\subsection{Emergent selection rules}

In the plot of relevant dynamic spin structure factors in Fig.~\ref{fig3}, 
the two branches of magnetic excitations in the quantum paramagnet are degenerate
due to the time reversal symmetry, while only one branch of magnetic excitations
in the $\text{FM}_{\text{xy}}$ and $\text{FQ}_{\text{xy}}$ states is visible. 
As we show below, the latter arises from the emergent selection rules.

We start with the $\text{FM}_{\text{xy}}$ state. 
For this state, the elastic neutron scattering is able to reveal 
an in-plane ferromagnetic dipolar order (e.g., along the $x$ direction
for the choice in the previous section). For the usual 
ferromagnet for spin-1/2 degrees of freedom, the dynamic spin
structure factor for spin components along the ferromagnetic ordering direction 
measures the two magnon continuum. 
For our effective spin-1 local moments that have a larger physical
Hilbert space, the microscopic interaction in the Hamiltonian 
could access the large Hilbert space at the linear order. This 
qualitatively explains the presence of coherent magnetic excitation 
in Fig.~\ref{fig3}(b). More specifically, within the flavor wave theory, 
the $S^x$ operator in the reciprocal space is written as 
\begin{eqnarray}
S_{\boldsymbol{k}}^{x}=i(1-4c^{2})\big(a_{\boldsymbol{k}2}^{\phantom{\dagger}}
-a_{-\boldsymbol{k}2}^{\dagger}\big)%
\end{eqnarray}
where the static piece for the ferromagnetic order has been ingored. 

The absence of one branch of magnetic excitation in Fig.~\ref{fig3}(b) 
is a consequence of the emergent selection rule.
Our Hamiltonian is invariant under the following $\mathbb{Z}_{2}$
symmetry operations generated by 
\begin{align}
\hat{G}_{x} & =e^{-i\pi\sum_{j}(S_{j}^{x}+1)},\label{eq:Gx}\\
\hat{G}_{y} & =e^{-i\pi\sum_{j}(S_{j}^{y}+1)}.\label{eq:Gy}
\end{align}
For the $\text{FM}_{\text{xy}}$ state, we have chosen the dipolar magnetization
along $\hat{x}$ direction such that the $\hat{G}_x$ symmetry is preserved.  
The flavor wave operators $a_{1}$ and $a_{2}$
(in Sec.~\ref{sec3c}) are odd and even under $\hat{G}_{x}$,
respectively. Therefore, the $S^{x}$ operation only excites the $a_{2}$
flavor with even parity while the $a_{1}$ band that has odd parity
is hidden in this measurement.

The same strategy can be applied to
the $\text{FQ}_{\text{xy}}$ state (we have chosen the quadrupolar director
along $x$ axis) but we need to consider the $\hat{G}_{y}$ symmetry.
Following the same argument, one can see that in the $S^{zz}$ channel
only odd-parity $a_{1}$ excitation can be measured in the spectrum. 
More explicitly, the $S^z$ operator is written as 
\begin{align}
S_{\boldsymbol{k}}^z=-i\big(a_{\boldsymbol{k}1}^{\phantom\dagger}-a_{-\boldsymbol{k}1}^{\dagger}\big)
\end{align}
in the flavor wave formulation of Sec.~\ref{sec3d}.

\section{Discussion}
\label{sec5}

\begin{table*}
\begin{tabular}{cccc}
\hline\hline
Systems and materials & Local low-lying states & Relevant condensation & Refs \\ 
\hline
Dimerized magnets & Singlet of two neighboring spins & Triplon from the triplets & 
Ref.~\onlinecite{dimer} \\
A-site spinel FeSc$_2$S$_4$ & Spin-orbital singlet with $e_g$ orbitals & Spin-orbital excited states & 
Refs.~\onlinecite{PhysRevLett.102.096406,PhysRevB.80.224409} \\
$4d^4/5d^4$ Mott insulators (Ca$_2$RuO$_4$) & Spin-orbital singlet with $t_{2g}$ orbitals &
Spin-orbital excitons &  Refs.~\onlinecite{PhysRevLett.111.197201,PhysRevB.90.035137,PhysRevLett.116.017203,PhysRevLett.119.067201}\\
$3d^8/4d^8$ Mott insulators & Spin-orbital singlet with $t_{2g}$ orbitals & Spin-orbital excitons & Ref.~\onlinecite{FYLee} \\
Magnets with weak crystal field gaps & Local low-lying crystal field state & Excited crystal field states & This work
\\
\hline\hline
\end{tabular}
\caption{List of physical contexts that support similar upper branch physics. 
``Relevant condensation'' in the table refers to either field driven or exchange 
interaction driven transition.}
\label{tab2}
\end{table*}

In summary, we have explored the physical consequences
of the weak crystal field splitting for which the standard approach
is insufficient to capture the low-enegy magnetic physics. For concreteness, we
study a model system with the weak crystal field level scheme on the
FCC lattice. Using both the Weiss mean-field method and the flavor
wave theory, we obtain the phase diagram of the model and reveal the effects
of the excited crystal field states. There are two different 
mechanisms from which various magnetic ordering happens, 
one is related to the exchange within
the excited levels and the other is related to the hybridization 
between the ground state level and the excited levels. The
FCC lattice may not be highly frustrated, and the states are mostly
ordered in our study. On more frustrated lattices, 
the collapses of the excited crystal field states may lead 
to more possibilities such as quantum spin liquid.

As for the physical relevance, we are currently not aware of many relevant 
physical systems that explicitly show the upper branch magnetism. The pyrochlore material
Tb$_{2}$Ti$_{2}$O$_{7}$ (and maybe other Tb-based pyrochlore magnets, e.g. 
Tb$_2$Sn$_2$O$_7$)~\cite{0953-8984-24-25-256003,PhysRevB.89.134410,tbcef15,tbcef07,molavian2007dynamically,gardner2003dynamic,PhysRevLett.82.1012,PhysRevLett.94.246402}
can be thought as a potential relevance. The crystal field
gap is not as large as other well-known pyrochlore magnets such 
as Yb$_2$Ti$_2$O$_7$ or Er$_2$Ti$_2$O$_7$.
For example, Ref.~\onlinecite{PhysRevLett.82.1012} 
actually measured a crystal field gap of 1.5meV
between the ground state doublet and the excited state doublet. 
Thus, the usage of the effective spin-1/2 local moment for the 
ground state doublet of the Tb$^{3+}$ ion
may not apply very well in certain cases. The magnetic entropy
of Tb$_2$Sn$_2$O$_7$ does increase beyond R$\ln 2$ as 
the temperature is increased beyond 4K~\cite{PhysRevLett.94.246402}. The inelastic neutron 
scattering measurement in Tb$_2$Ti$_2$O$_7$ shows a clear 
dispersion for the excited doublets with a renormalized
gap~\cite{refId0}. All these phenomena suggest the importance of the upper branch 
physics. It would be interesting in the future to actually
suppress the crystal field gap and drive the system
to magnetic orders by collapsing the excited levels. 
In fact, Tb$_2$Sn$_2$O$_7$ experiences a magnetic ordering transition 
around 1K~\cite{PhysRevLett.94.246402}. 
The actual modeling of the magnetic physics should be in terms 
of an effective ${J=3/2}$ local moment that takes care of both
the ground state doublet and the first excited state doublet,
and the interaction would naturally be a $\Gamma$-matrix model.

For the $4d/5d$ magnets such as iridates and others, the often used 
description is in terms of the spin-orbit-entangled local moment 
$J$~\cite{WCKB}. It should certainly be the case if the local ground state $J$ state
is well separated from the excited $J$ states. The $4d$/$5d$ orbitals, however,
are very extended. Very often, the hybridization of the superexchange 
involving the upper excited $J$ states may be not that small compared to the 
local energy gap due to the spin-orbit coupling. This piece of physics 
has been nicely invoked by G. Khaliullin in Ref.~\onlinecite{PhysRevLett.111.197201} 
for the $4d^4/5d^4$ magnets, 
where he described
the physics as the singlet-triplet condensation to make the analogy 
with the triplon condensation in the dimerized magnets. 
Here we think this physics is not restricted to the $4d^4/5d^4$ magnets
whose local ground state would be a trivial spin-orbit singlet with ${J=0}$,
but extends broadly to many other non-singlet spin-orbit-coupled magnets
(as long as the spin-orbit-coupling induced local gap is not large).

In Table.~\ref{tab2}, we list the relevant systems/materials that could 
share a similar physics as the upper branch physics in this work. Our result
simply provides a new member to this list of ``triplon''-like physics. 
We expect the universal physics like the Higgs mode (or amplitude mode)
could also emerge in the relevant materials of our work 
where the conendensation or criticality is from the collapse of the excited 
crystal field states.

\section{Acknowledgments}

This work is supported by the ministry of science and technology of China 
with Grant No.~2016YFA0301001, 2016YFA0300500, 
the start-up fund and the first-class University construction fund of 
Fudan University, and the thousand-youth-talent program of China.

\bibliography{refs}

\end{document}